\documentclass[a4paper]{article}

\usepackage[dvipsnames]{xcolor}
\usepackage{INTERSPEECH2020}
\usepackage{subcaption}

\def\L{{\cal L}}
\DeclareMathOperator{\sinc}{sinc} 

\DeclareMathOperator{\blstmp}{BLSTMP}
\DeclareMathOperator{\lstm}{LSTM}
\DeclareMathOperator{\softmax}{softmax}
\DeclareMathOperator{\linear}{Linear}
\DeclareMathOperator{\logc}{LogCompression}

\DeclareMathOperator{\DConvBlock}{DConvBlock}
\DeclareMathOperator{\SincBlock}{SincBlock}

\title{Lightweight End-to-End Speech Recognition from Raw Audio Data Using Sinc-Convolutions}
\name{Ludwig K{\"u}rzinger$^1$, Nicolas Lindae$^1$, Palle Klewitz$^1$, Gerhard Rigoll$^1$}
\address{
  $^1$Technische Universit{\"a}t M{\"u}nchen}
\email{ludwig.kuerzinger@tum.de, nicolas.lindae@tum.de, palle.klewitz@tum.de}

\newcommand{\tabheader}[1]{\multicolumn{1}{c}{\textbf{#1}}}

\begin{document}

\maketitle
\begin{abstract}

Many end-to-end Automatic Speech Recognition (ASR) systems still rely on pre-processed frequency-domain features that are handcrafted to emulate the human hearing.
Our work is motivated by recent advances in integrated learnable feature extraction.
For this, we propose Lightweight Sinc-Convolutions (LSC) that integrate Sinc-convolutions with depthwise convolutions as a low-parameter machine-learnable feature extraction for end-to-end ASR systems.

We integrated LSC into the hybrid CTC/attention architecture for evaluation.
The resulting end-to-end model shows smooth convergence behaviour that is further improved by applying SpecAugment in time-domain. 
We also discuss filter-level improvements, such as using log-compression as activation function.
Our model achieves a word error rate of $10.7\%$ on the TEDlium v2 test dataset,
surpassing the corresponding architecture with log-mel filterbank features by an absolute $1.9\%$,
but only has $21\%$ of its model size.
\end{abstract}


\section{Introduction}
\label{sec:intro}

End-to-end Automatic Speech Recognition (ASR) systems directly infer tokens (letters or word pieces); sentence transcription is done in a single stage without intermediate representations, e.g. \cite{WatanabeEtAl17,FacebookRaw1,hannun2014deep,RawAudioWWAmazon}.
This structure is desirable, as transcriptions can be directly used as a training objective without having to rely on hard-coded feature extraction or intermediary objectives, such as separately trained acoustic models.
However, one may argue that many architectures are not yet fully end-to-end as the encoder still requires preprocessed audio features in the form of log-mel filterbanks.
Recent advances using parametrizable Sinc-convolutions enable the integration of the feature extraction as learnable part into the neural network~\cite{SincNet1, SincNet2,SincNet3}.

In this work, we propose a {lightweight} machine-learnable feature extraction integrated into the ASR neural network to directly classify sentences from raw audio.
For this, we extend the established hybrid CTC/Attention system~\cite{WatanabeEtAl17} architecture, that combines two main techniques for end-to-end ASR:
Neural networks trained with Connectionist Temporal Classification (CTC~\cite{GravesEtAl06}) calculate the probability of each token at a given time step in the input sequence.
Attention-based encoder-decoder architectures~\cite{ChanEtAl16} are trained as sequence-generative models to transform the input sequence into a transcribed token sequence.

We term our proposed extension Lightweight Sinc-Convolutions (LSC); it combines Sinc-convolutions with depthwise convolutions.
A similar network was originally proposed for lightweight keyword spotting~\cite{mittermaier2020small} optimized towards low power consumption.
Lightweight architectures with less parameters consume less energy on battery-driven devices, since memory access causes far more power consumption as the computation itself~\cite{MemVsOps}.

\textbf{Our contributions:}
\begin{itemize}
	\item We propose Lightweight Sinc-Convolutions (LSC): A low-parameter front-end that can be integrated into end-to-end ASR architectures to decode from raw audio.
	\item For performance evaluation, we integrated LSC into a parameter-reduced model of the established hybrid CTC/attention ASR system.
	\item In combination with a high-capacity language model, our reduced model achieves a competitive word error rate of $10.7\%$ on the TEDlium v2 test set.
\end{itemize}


\section{Related Work}
\label{sec:related}

Recent approaches in end-to-end speech recognition started to outperform conventional hybrid DNN/HMM speech recognition~\cite{WatanabeEtAl17,FacebookRaw1,hannun2014deep,RawAudioWWAmazon,WatanabeEtAl18}.
However, most of these end-to-end systems, such as~\cite{WatanabeEtAl17}, still rely on preprocessed log-mel F-bank features.
A common approach is also to classify using similar methods as in image recognition.
For example, Hannun et al.~\cite{hannun2014deep} train an end-to-end recurrent network on spectrogram features.
But only few architectures directly classify on raw audio~\cite{RawAudioWWAmazon}.

There are two dominating techniques for end-to-end speech recognition:
(1) Connectionist Temporal Classification (CTC~\cite{GravesEtAl06}) carries the concept of hidden Markov states over to end-to-end neural networks as training loss for sequence-classification networks.
Neural networks trained with CTC loss calculate the posterior probability of each letter at a given time step in the input sequence.
(2) Attention-based encoder-decoder architectures~\cite{ChanEtAl16} are trained as autoregressive sequence-generative models.
The encoder transforms the input sequence into a latent representation.
From this, the decoder generates the sentence transcription.
Attention-based sequence transcription was proposed in the field of machine language translation in~\cite{BahdanauEtAl14} and applied to speech recognition in~\cite{ChanEtAl16}.
Our work builds on the hybrid CTC/Attention ASR system that combines CTC with the location-aware attention mechanism, as proposed in~\cite{ChorowskiEtAl15}.
Sentence transcription is performed with the help of a RNN language model (RNNLM) integrated into decoding process using {shallow fusion}~\cite{GulcehreEtAl15}.
Detailed descriptions of the hybrid CTC/attention ASR architecture can be found in~\cite{WatanabeEtAl17,WatanabeEtAl18}.

Our work is motivated by recent developments in speech and speaker recognition on learnable feature extraction that can be integrated into an end-to-end trainable neural net~\cite{SincNet1,SincNet2,SincNet3,mittermaier2020small,latif2020deep,Won2020HarmonicFilt,SincNet4,SincNet5}.
Ravanelli et al. demonstrated that parametrizable convolutions with a filter structure have better convergence behavior than CNNs alone on raw audio data and proposed the SincNet architecture~\cite{SincNet1}
that was first used in speaker recognition~\cite{SincNet1, SincNet2} and then phoneme recognition~\cite{SincNet3}.
E2E-SincNet~\cite{SincNet4}, proposed by Parcollet et al., integrates the SincNet architecture into the hybrid CTC/Attention architecture~\cite{WatanabeEtAl17}.

Sinc-convolutions were also successfully applied to low-parameter keyword spotting~\cite{mittermaier2020small} to optimize power consumption in battery-driven devices.
The model proposed by Mittermaier et al. employs depthwise separable convolutions~\cite{DSConvforTranslation, Xception} on top of Sinc-convolutions.
Depthwise separable convolutions are useful in low-parameter scenarios~\cite{Mobilnets,HelloEdge}.
A similar approach to reduce the number of parameters in convolutions are lightweight convolutions~\cite{wu2019pay} that are based on depthwise convolutions.
In our proposed method, we employ depthwise convolutions that omit the pointwise convolution of depthwise separable convolutions~\cite{DSConvforTranslation}. 


\section{Architecture}
\label{sec:architecture}

This section describes the proposed feature extraction block as our main contribution and, in a second part, its integration into an existing end-to-end ASR system.
This architecture directly infers token and sequence probabilities from raw audio, without any previously extracted log-mel features.

\subsection{Front-End: Lightweight Sinc-Convolutions (LSC)}

\begin{figure}[b!]
  \centering
  \includegraphics[width=0.90\columnwidth]{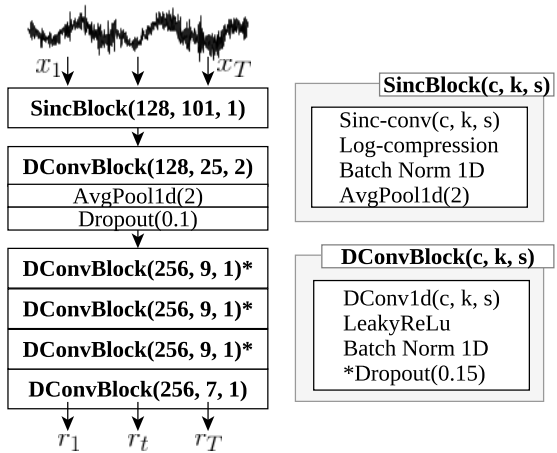}
  \caption{The proposed Lightweight Sinc-Convolutions (LSC).
  The raw audio stream is partitioned into frames.
  A parametrizable Sinc-convolution and multiple layers of DConvs are then applied for lightweight extractions of features.}
  \label{fig:frontend}
\end{figure}

To obtain higher-level latent representations from a sequence of raw audio frames $X=x_{1:T}$, we build upon one layer of parametrized Sinc-convolution and several layers of depthwise convolutions (DConv).
These layers extract filter features $R$ from our input, replacing pre-calculated log-mel F-bank features that are commonly used in end-to-end ASR systems:
\begin{equation}
R=r_{1:T} = [\DConvBlock]^{5}(\SincBlock(x_{1:T}))
\end{equation}
We term this architecture Lightweight Sinc-Convolutions (LSC), as shown in Fig.~\ref{fig:frontend}.
Based on the Sinc function $\sinc (x)=\sin (x)/x$, the Sinc-convolution is constructed from the convolutional kernel
\begin{equation}
g[n,f_1,f_2]= 2 f_2 \sinc (2\pi f_2 n) - 2 f_1 \sinc (2\pi f_1 n),
\end{equation}
that acts as a bandpass filter in the spectral domain with $f_1$ and $f_2$ as the cutoff frequencies.
Unlike with pre-computed filterbank features, the above formula can be integrated into a neural network as a layer that is differentiable w.r.t. the cutoff frequencies.
To better approximate an ideal Sinc filter, the Hamming window $w[n] = 0.54 - 0.46 \cos ( {2 \pi n}/{L} )$ is multiplied with the L weights in the Sinc-convolution kernel.
The cutoff frequencies of a bandpass filter are derived from the learnable parameters $w_1$ and $w_2$~\cite{SincNet1}: 
\begin{align}
f_1 &= |w_1| \\
f_2 &= |w_1| + |w_2 - w_1|
\end{align}
For non-linearity, we use log-compression as activation function of this layer, defined as $\logc(x) = \log(|x|+1)$.
In comparison with the ReLU activation function as used in SincNet~\cite{SincNet1}, log-compression drops less information and can improve the classification performance on raw audio~\cite{FacebookRaw1,mittermaier2020small,MittermaierMA,FacebookRaw2}.

The Sinc-convolutional layer is followed by multiple depthwise convolutions (DConvs) for extraction of higher-level latent representations.
DConvs use separate kernels for each input channel, reducing the number of parameters of a DConv layer with $c_{in}$ input and $c_{out} = n\cdot c_{in}, n\in\mathbb{N}$ output channels and a kernel of size $k$  from $c_{in}\cdot c_{out}\cdot k$ to $c_{out}\cdot k$.
They extract short-time context along the time dimension, however, limited to the data within one frame.
This part of the architecture acts as coupling layer between the Sinc-convoluted signal and the ASR back-end network.
LSC is heavily inspired by the lightweight keyword spotting architecture by Mittermaier et al.~\cite{mittermaier2020small} using depthwise {separable} convolutions.
Our adaption mainly omits the inter-frame pointwise convolution in order to further reduce the number of parameters.
DConvs contribute more to the extraction of high-level features than pointwise convolutions, while making up only a small fraction of the learnable parameters~\cite{guo2019depthwise}.

\subsection{Back-End ASR Architecture}
\begin{figure}[ht!]
  \centering
  \includegraphics[width=0.90\columnwidth]{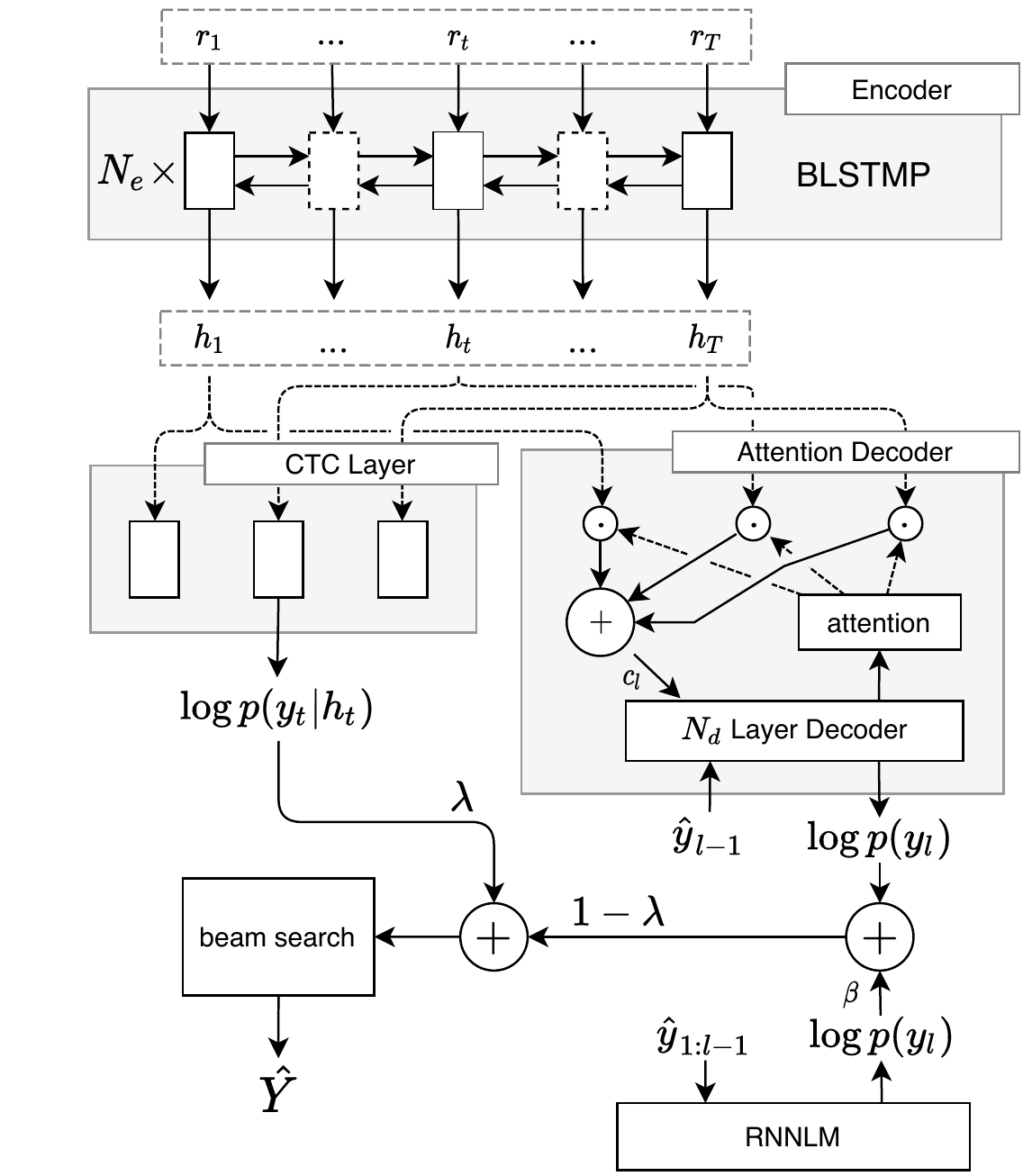}
  \caption{ASR Back-end: Features extracted by the front-end are passed through a BLSTMP encoder. Encoder states are used by a CTC and an attention decoder to decode the most probable sequence.
  The decoding process is guided by a beam search.}
  \label{fig:architecture}
\end{figure}
The back-end architecture transforms the extracted features provided by the LSC front-end into a sentence, as shown in Fig.~\ref{fig:architecture}.
The first part, the encoder, consists of BLSTM layers with projection layers (BLSTMP) as proposed in \cite{WatanabeEtAl17,WatanabeEtAl18}.
It maps the sequence of latent representations $R = r_{1:T}$ to a sequence of high-level representations, i.e., $H = h_{1:T} = \blstmp(x_{1:T})$.

On top of the encoder, one layer maps the hidden state sequence $h_{1:T}$ to the predicted output sequence of tokens $\hat{y}_{1:L_1}^\mathrm{CTC}$ of shorter or equal length $L_1$.
This layer is trained with Connectionist Temporal Classification (CTC) loss.
For this, we first use a linear layer with softmax activation to map the hidden state sequence to a sequence of token probabilities, where the list of tokens has been extended with a blank token: $\hat{y}'_{1:T} = \softmax(\linear(h_{1:T}))$.
Each $\hat{y}'_t$ is a probability distribution over the tokens for timestep $t$. 
From these, the probability $p_\mathrm{CTC}(Y|X)$ of our target sequence $Y = y_{1:L}$ is calculated by summing up the probabilities of all possible paths through $\hat{y}'_{1:T}$ leading to the target sequence, where we remove duplicate symbols that aren't separated by a blank token.
The CTC loss is the cross entropy of that probability: $\L_\mathrm{CTC} = -\log p_\mathrm{CTC}(Y|X) $.
For beam search decoding, the prediction $\hat{y}_{1:L_1}^\mathrm{CTC}$ for the most probable sequence is obtained using a modified forward algorithm~\cite{WatanabeEtAl17}.

The hybrid CTC/Attention architecture incorporates a LSTM based decoder network that utilizes location-aware attention as a method of attending arbitrary encoder states and thereby eliminating the bottleneck between encoder and decoder.
At step $l$, location aware attention retrieves arbitrary encoder states by computing an importance score $score(h_t, q_{l_,t})$ for every encoder state $h_t$, given that state, the previous decoder state $q_{l-1}$, as well as the previous attention distribution $a_{l-1}$:
\begin{align}
\mathrm{score}(h_t, q_{l-1}) =  &g^\mathrm{T} \cdot \tanh\big(\linear([q_{l-1}, h_{t}, K * a_{l-1, t}])\big) \nonumber
\end{align}
Here, $*$ denotes the convolution operator with a kernel $K$. 
To reduce the attention score to a scalar, the vector outputs of each linear layer are added and the dot product with a learned vector $g$ is computed.
Attention scores are normalized by a softmax:
\begin{equation}
    a_{l,t} = \frac{\exp(\mathrm{score}(h_t, q_{l-1}))}{ \sum_{1 \leq j \leq T} \exp( \mathrm{score}(h_j, q_{l-1}))}
\end{equation}
This attention weight is then multiplied with its corresponding encoder state and the attended state vectors are reduced into the context vector $c_l =\sum_{1\leq t \leq T} a_{l,t} \cdot h_t$.
The decoder network then autoregressively produces posterior token probabilities given the embedding of the previous hypothesized token $\hat{y}_{l-1}$ concatenated with the attention context:
\begin{align}
    q_l &= [\lstm]^{N_\text{d}}([c_t, \mathrm{Embedding}(\hat{y}_{l-1})], q_{l-1}) \\[.5em]
    p_\mathrm{Att}(y_l)& = \softmax(\linear(q_l))
\end{align}

The loss for the attention decoder is also calculated using cross entropy: $\L_\mathrm{Att} = -\log p_\mathrm{Att}(Y|X)$.
To improve the grammatical correctness of predictions, we incorporate an additional pre-trained language model into our system using shallow fusion as introduced by Gulcehre et al.~\cite{GulcehreEtAl15}. 
Shallow fusion combines the log probabilities of the attention model with the log probabilities of the language model using a weight $\beta$, i.e., $\log{p(\hat{y}_l)} = \log{p_\mathrm{Att}(\hat{y}_l)} + \beta \cdot  \log{p_\mathrm{LM}(\hat{y}_l)}$.

Finally, for a hybrid CTC/Attention architecture, we combine the CTC loss with the attention loss by weighing them in a multi-objective loss function $\mathcal{L} = (1-\lambda)\L_\mathrm{Att} + \lambda\L_\mathrm{CTC}$.
Likewise, to predict a sequence from our input we sum up the probabilities from the attention network, from CTC and additionally from the language model and estimate the most probable target sequence from that using beam search, as shown at the bottom of Fig.~\ref{fig:architecture}.

\section{Experiments}
\label{sec:experiments}

We evaluate our architecture on the freely available TEDlium v2 corpus~\cite{RousseauEtAl14} with more than 200 hours of speech.
It is a complex ASR task widely used for speech recognition.
The raw audio stream is split into frames of $25$ ms of raw audio, with two consecutive frames overlapping by $15$ ms, resulting in a stride of $10$ ms\footnote{
Frame parameters were chosen according to the standard setting for MFCC features in the kaldi toolkit~\cite{povey2011kaldi}.}.
As the audio was recorded with a sampling frequency of 16kHz, this means our frames consist of $400$ samples each, the overlapping section of $240$ samples.
We trained our network using the Adadelta optimizer ($\rho=0.95$, $\epsilon=10^{-8}$) for 22 epochs before early stopping ended training, and then chose the model with the highest validation accuracy for evaluation.
The Sinc convolution consists of 128 Sinc filters initialized with mel filterbank weights.
For our network's encoder, we chose to use four BLSTM layers with projection neurons of size 512. The decoder is a single LSTM layer of size 512 in combination with an attention layer also of size 512.
We chose 0.5 as our CTC/Attention weight and trained with 500 unigram units as tokens, created by SentencePiece~\cite{kudo2018sentencepiece}.

As additional data augmentation, we add spectral augmentation~\cite{Park2019} to the training procedure.
Spectral augmentation introduces two augmentations techniques:
Time warping distorts part of the input features, and the dropout-like masking is applied to blocks in both feature and time dimension.
Although this technique is meant to be used on already extracted filter features, we found it to improve ASR performance.
We used the standard configuration as provided by the ESPnet toolkit, although it is sensible to further investigate and adjust these values in the future.
For decoding, we used a pre-trained language model provided by ESPnet.
The language model was integrated with a fusion weight of $\beta=0.5$, the CTC weight was chosen to be $\lambda=0.4$.

\section{Results and Discussion}
\label{sec:discussion}

As a baseline model to compare to, we use the hybrid CTC/attention ASR as proposed in~\cite{WatanabeEtAl17,WatanabeEtAl18}.
It also uses the location-aware attention mechanism, but classifies from log-mel FBank features.
We compare our model performance to the best published VGG-BLSTMP result~\cite{WatanabeEtAl18}\footnote{Results obtained from the best VGG-BLSTMP model as listed in the ESPnet github repository at commit hash f36bbee}.
An overview that shows parameters of our architecture compared to this model from ESPnet is given in Table \ref{tab:features}.
Our network is smaller in the number of learnable parameters compared to VGG-BLSTMP.
We managed to create a front-end with fewer parameters due to Sinc and depthwise convolutions:
LSC uses only $16$k parameters, whereas the VGG layers learn $259$k parameters.
Our low parameter count can be attributed to depthwise convolutions.
In the proposed parameter configuration as seen in Fig.~\ref{fig:frontend}, adding the pointwise convolution in the DConv blocks would result in additional $240$k parameters.
We were also able to use thinner back-end layers of size $512$ (instead of $1024$ used in the configuration of the VGG-BLSTMP model), resulting in only $22.3$m parameters instead of $106$m.
The CTC layer requires an architecture that provides a short-term context;
here, it's provided by the BLSTM in the encoder, so that frame-wise 1D filter convolutions can be used without loosing inter-frame information. 

\begin{table}[ht!]
\centering
\caption{Architecture comparison between the hybrid CTC/attention VGG-BLSTMP model~\cite{WatanabeEtAl17} and our LSC+BLSTMP architecture.}
\label{tab:features}
\begin{tabular}{lll}
\hline
          & \tabheader{VGG+BLSTMP}  & \tabheader{LSC+BLSTMP} \\
          & \tabheader{(ESPnet)}  & \tabheader{(Ours)} \\
\hline
Type & F-bank+pitch          & \textbf{Sinc-Convs}  \\
Window        & Hann~\cite{oppenheim1999discrete}         & Hamming     \\
Frame size     & $25$ ms     &   $25$ ms   \\
Frame shift     & $10$ ms    &   $10$ ms   \\
Freq. bins     & 80+3                &  128    \\
Coupling        & VGGnet              & \textbf{DConvs}   \\
Output dim.        & $2688$       &  $256$ \\
Size front-end        & $259$k        &  $\mathbf{16k}$ \\ 
Size back-end        & $106$m        &  $\mathbf{22.3m}$ \\ 
\hline
\end{tabular}
\end{table}

\begin{table}[ht!]
    \caption{Comparison of perfomance on the TEDlium 2 dataset~\cite{RousseauEtAl14}.
    Listed are the best VGG-BLSTMP WER from ESPnet~\cite{WatanabeEtAl18}, of a VGG-BLSTMP with a to our model comparable network size; and the variations of our architecture with LSC.}
    \centering
    \begin{tabular}{l c c r r}
        \hline
         \tabheader{Model} & \tabheader{Fea.} & \tabheader{LM} & \tabheader{Dev} & \tabheader{Test} \\
        \hline
         VGG-BLSTMP~\cite{WatanabeEtAl17} & FBANK & \checkmark & 12.8 & 12.6 \\
         VGG-BLSTMP (small) & FBANK & \checkmark & 19.8 & 18.6 \\
         LSC & RAW & - & $15.1$ & $15.7$ \\
         LSC+SpecAug & RAW & - & $13.5$ & $13.4$ \\
         LSC & RAW & \checkmark & $11.4$ & $11.6$ \\
         LSC+SpecAug (ReLU) & RAW & \checkmark & $11.0$ & $10.9$ \\
         LSC+SpecAug (logCpr) & RAW & \checkmark & $\mathbf{10.7}$ & $\mathbf{10.7}$ \\
        \hline
    \end{tabular}
    \label{tab:resultsted}
\end{table}

Despite its smaller size, the model still achieves a better WER performance on the TEDlium v2 test set.
By inspection, a smooth asymptote-like of model loss convergence can be observed during training on this dataset.
Table \ref{tab:resultsted} shows performance comparisons of our architecture on the TEDlium 2 dataset with the VGG-BLSTMP model.
We achieve a $0.9\%$ lower WER of $11.7\%$ on the test set when decoding with a language model of similar size ($8.9$m compared to $7.4$m parameters).
When using the large language model used for ESPnet's transformer network, we can further reduce the WER by $1\%$ to $10.7\%$, however, the parameter count of the larger LM increases to $139$m parameters.
Although we use a large RNNLM for evaluation, a more lightweight LM can be implemented depending on application scenario.
To provide a more distinctive and fair comparison, we also list a parameter-reduced VGG-BLSTMP model with only $19$m parameters that is comparable in model size to our evaluated model.
This model however only achieved a WER of $18.6\%$~\cite{Karita2019}.

Training with spectral augmentation also improved the network's performance, despite not fine tuning the configuration for raw audio data as input:
Spectral augmentation improved the WER by $2.3\%$ and $0.9\%$ when decoding without and with the large language model.
To evaluate log-compression as alternative activation, we also tested one model instance using ReLU as activation function in the SincBlock (as in Fig.~\ref{fig:frontend}).
With ReLU, we achieved slightly worse WERs of $11.0\%/10.9\%$ for the dev/test set, confirming the results from Mittermaier et al.~\cite{mittermaier2020small, MittermaierMA}.
This indicates that log-compression is benefical to the extraction of filter features from raw audio.

\begin{figure}[t!]
    \centering
    \includegraphics[width=0.88\columnwidth]{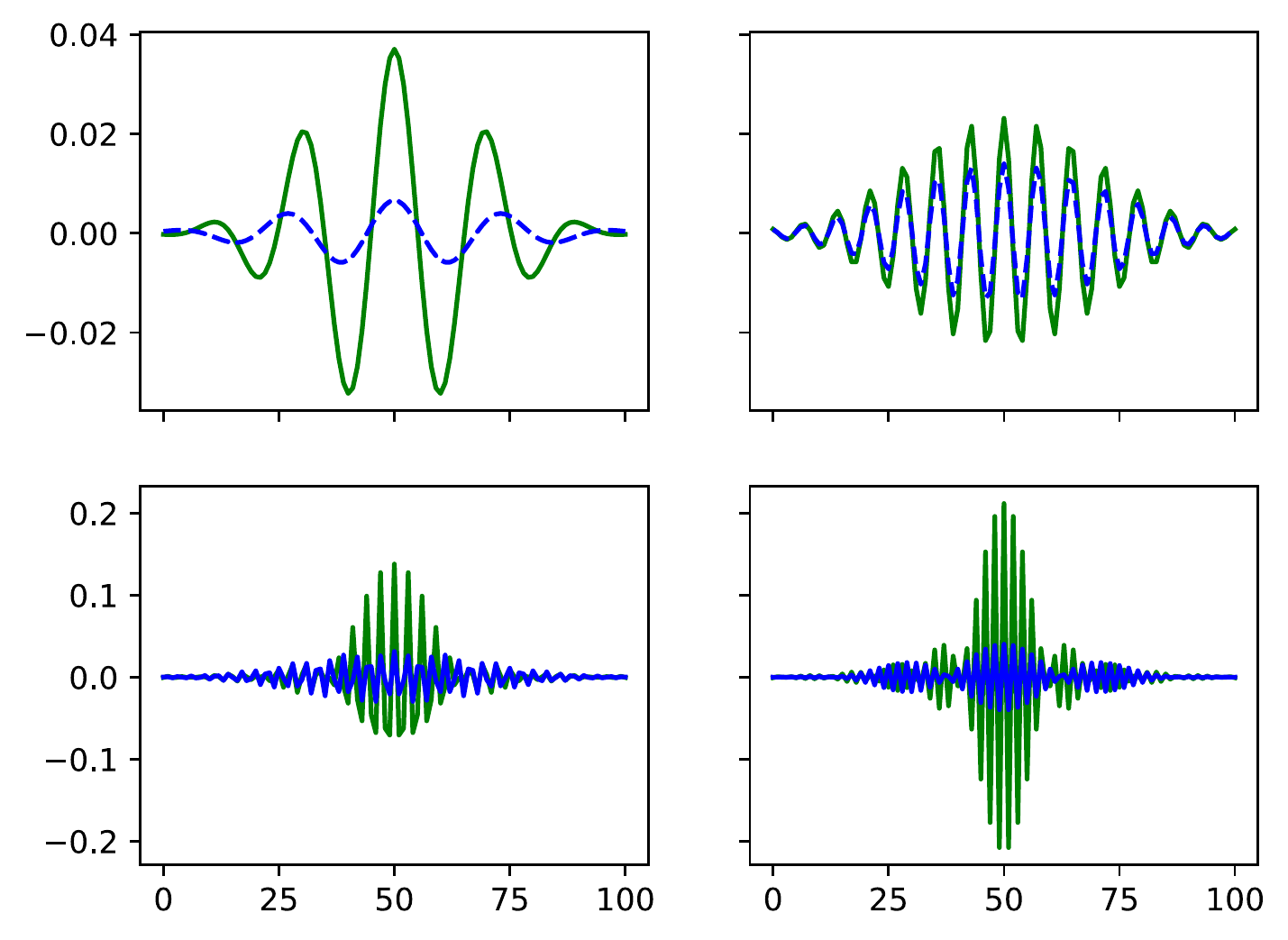}
    \caption{Four exemplary Sinc-convolution kernels.
    The blue dashed line indicates the corresponding mel-scale filter.
    Their center frequencies (from left to right, top to bottom) are $790$Hz, $2.2$kHz, $5.4$kHz, $7.9$kHz.}
    \label{fig:sincfilterex2}
\end{figure}

\begin{figure}[htb!]
  \centering
  \includegraphics[width=0.88\columnwidth]{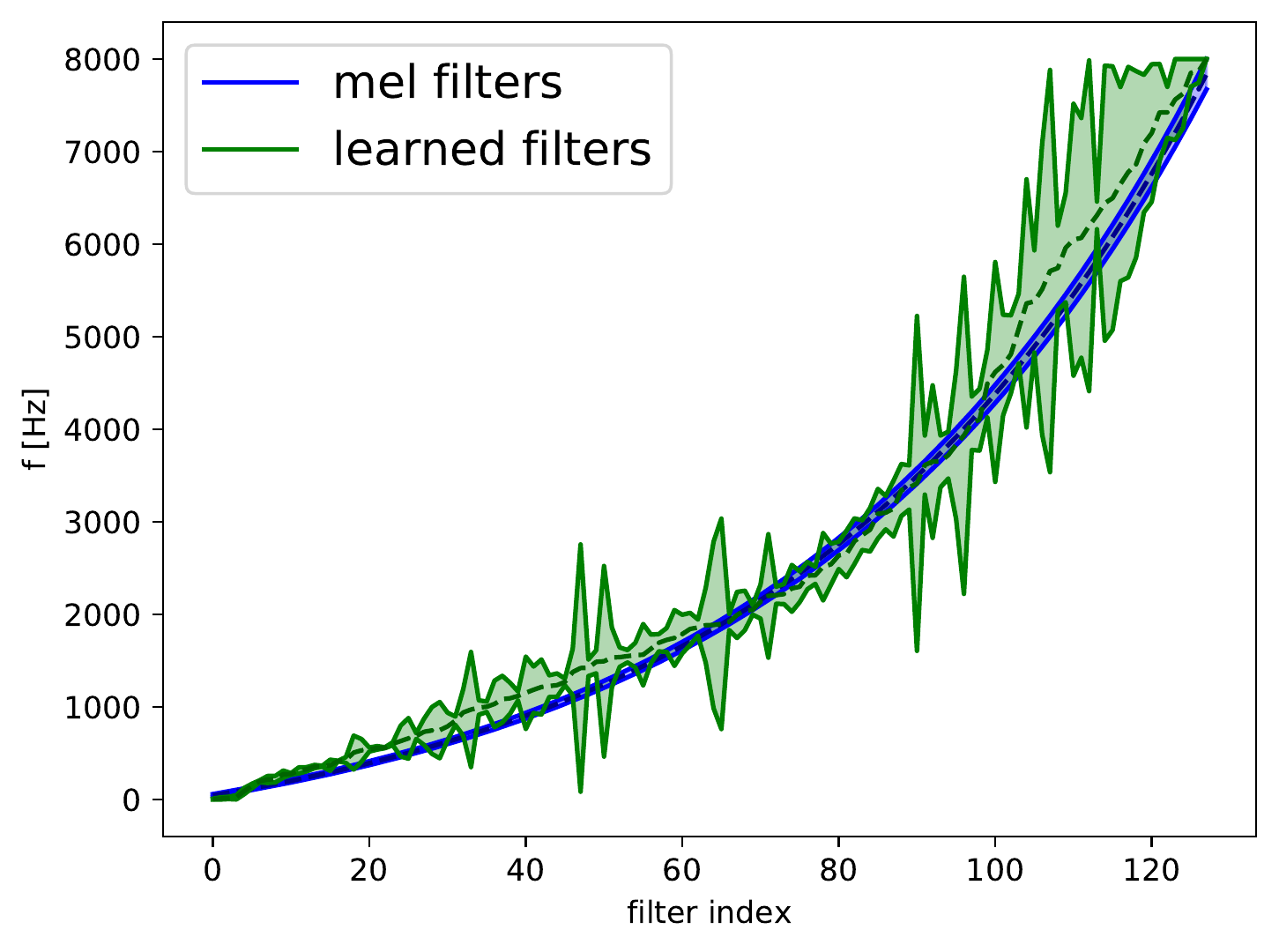}
  \caption{Learned Sinc-convolution filters visualized by plotting the lower and upper bounds of the filters.}
  \label{fig:filtergraph}
\end{figure}

Fig.~\ref{fig:sincfilterex2} shows four learned Sinc-convolution filters of the LSC, and Fig.~\ref{fig:filtergraph} visualizes the learned filters by sorting them and plotting their upper and lower bounds.
In those filters, a trend towards a higher amplitude is noticeable, also towards a wider band pass in the spectral domain.
Notably, one of the Sinc filters converged to pass through the entire raw audio signal, indicating an inclination of the network to directly process the raw data.

\section{Conclusions}
\label{sec:conclusion}
Many  end-to-end  ASR systems still rely on pre-processed f-Bank features.
Our work extended an existing ASR system with a front-end to classify directly from raw audio that we termed lightweight Sinc-convolutions (LSC).
Its network architecture is carefully chosen to be lightweight, i.e. using less parameters, which is an advantage for speech recognition in battery-driven devices.
We discussed further improvements to this architecture, such as log-compression as activation function and spectral augmentation for time-based data augmentation.
Our final model achieves a word error rate of $10.7\%$ when combined with a large RNN language model, an improvement of absolute $1.9\%$ over the best reported model of the corresponding f-Bank architecture, but only has $21\%$ of its model size.

\bibliographystyle{IEEEtran}

\bibliography{mybib}

\end{document}